# A comparison of 7 Tesla MR spectroscopic imaging and 3 Tesla MR fingerprinting for tumor localization in glioma patients

Philipp Lazen[1,2,3], Pedro Lima Cardoso[1], Sukrit Sharma[1], Cornelius Cadrien[1,2], Thomas Roetzer-Pejrimovsky[4], Julia Furtner[5,6], Bernhard Strasser[1], Lukas Hingerl[1], Alexandra Lipka[1], Matthias Preusser[7], Wolfgang Marik[5], Wolfgang Bogner[1,3], Georg Widhalm[2], Karl Rössler[2,3], Siegfried Trattnig*[1,3,8], Gilbert Hangel[1,2,3]

[1]High-field MR Center, Department of Biomedical Imaging and Image-guided Therapy, Medical University of Vienna, Vienna, Austria
[2]Department for Neurosurgery, Medical University of Vienna, Vienna, Austria
[3]Christian Doppler Laboratory for MR Imaging Biomarkers, Vienna, Austria
[4]Division of Neuropathology and Neurochemistry, Department of Neurology, Medical University of Vienna, Vienna, Austria
[5]Division of Neuroradiology and Musculoskeletal Radiology, Department of Biomedical Imaging and Image-guided Therapy, Medical University of Vienna, Vienna, Austria
[6]Research Center for Medical Image Analysis and Artificial Intelligence (MIAAI), Danube Private University, Krems, Austria
[7]Division of Oncology, Department of Internal Medicine I, Medical University of Vienna, Vienna, Austria
[8]Institute for Clinical Molecular MRI, Karl Landsteiner Society, St. Pölten, Austria

*corresponding author (phone: +43 1 40400 64600, mail: siegfried.trattnig@meduniwien.ac.at)

# Abstract

This paper investigates the correlation between magnetic resonance spectroscopic imaging (MRSI) and magnetic resonance fingerprinting (MRF) in glioma patients by comparing neuro-oncological markers obtained from MRSI to T1/T2 maps from MRF.

Data from 12 consenting patients with gliomas were analyzed by defining hotspots for T1, T2 and various metabolic ratios, and comparing them using Sørensen-Dice Similarity Coefficients (DSCs) and the distances between their centers of intensity (COIDs).

Median DSCs between MRF and the tumor segmentation were 0.73 (T1) and 0.79 (T2). The DSCs between MRSI and MRF were highest for Gln/tNAA (T1: 0.75, T2: 0.80, tumor: 0.78), followed by Gly/tNAA (T1: 0.57, T2: 0.62, tumor: 0.54) and tCho/tNAA (T1: 0.61, T2: 0.58, tumor: 0.45). The median values in the tumor hotspot were T1=1724 ms, T2=86 ms, Gln/tNAA=0.61, Gly/tNAA=0.28, Ins/tNAA=1.15, and tCho/tNAA=0.48, and, in the peritumoral region, were T1=1756 ms, T2=102ms, Gln/tNAA=0.38, Gly/tNAA=0.20, Ins/tNAA=1.06, and tCho/tNAA=0.38, and, in the NAWM, were T1=950 ms, T2=43 ms, Gln/tNAA=0.16, Gly/tNAA=0.07, Ins/tNAA=0.54, and tCho/tNAA=0.20.

The results of this study constitute the first comparison of 7T MRSI and 3T MRF, showing a good correspondence between these methods.

## Key Points
- 7T MR spectroscopic imaging (MRSI) and 3T MR Fingerprinting (MRF) are two modern imaging methods which can complement MRI in the imaging of gliomas.
- Hotspots of MRSI's metabolic ratio glutamine (Gln) to total N-acetylaspartate (tNAA) and MRF's T2 map correspond very well to each other and to a radiologist's tumor segmentation.
- This work reinforces our hypothesis that the ratios of Gln/tNAA and Glycine/tNAA are promising tumor markers.

## Abbreviations
- AD — Acquisition delay
- Cho — Choline
- COID — Center of intensity differences
- Cr — Creatine
- CRT — Concentric ring trajectories
- DSC — Sørensen-Dice similarity coefficient
- FID — Free induction decay
- FISP — Fast Imaging with Steady-State Precession
- FLAIR — Fluid-attenuated inversion recovery
- FOV — Field of view
- Gln — Glutamine
- Glu — Glutamate
- Gly — Glycine
- Ins — Myo-inositol
- MP2RAGE — Magnetization-prepared 2 Rapid Gradient-Echo
- MRF — Magnetic resonance fingerprinting
- MRI — Magnetic resonance imaging
- MRSI — Magnetic resonance spectroscopic imaging
- NAA — N-acetylaspartate
- NAWM — Normal-appearing white matter

- PET       Positron emission tomography
- PT        Peritumoral segmentation
- ROI       Region of interest
- SNR       Signal to noise ratio
- TA        Acquisition time
- tCho      Total choline
- tCr       Total creatine
- TE        Echo time
- tNAA      Total N-acetylaspartate
- TR        Repetition time
- TU        Tumor segmentation

# Introduction

Over the last few decades, different approaches to magnetic resonance imaging (MRI) to produce different contrasts and images have been developed, including 3T-MR fingerprinting (MRF) and 7T high resolution MR spectroscopic imaging (MRSI), which aim to accumulate more specific information about brain tumors than conventional T1/T2-weighted MR imaging.

**Magnetic Resonance Spectroscopic Imaging**

MRSI provides metabolic information beyond contrast-enhanced T1/T2 MRI. The methodology visualizes different neurochemical concentrations without the need for contrast agents, and is thus a powerful tool in the investigation of diseases that influence metabolite and neurotransmitter distributions in the brain, such as gliomas. Notably, certain metabolites, such as N-acetylaspartate (NAA), creatine (Cr), choline (Cho), glutamine (Gln), glycine (Gly), and myo-inositol (Ins), are well suited as neuro-oncological markers due to the differences in concentration between tumor and healthy brain tissue and because of their stability in spectroscopic imaging, based on the accumulated experience of 7T MRSI in gliomas [1,2].

Our MRSI approach acquires free induction decay (FID) signals, following concentric ring trajectories (CRTs) in k-space [3]. Apart from the method's high sensitivity, one of the main advantages is its time efficiency. CRT-FID-MRSI can achieve high-resolution metabolic maps with a 64x64x39 matrix that covers the whole brain using an isotropic voxel size of 3.4 mm in 15 minutes, which presents a significant improvement compared to clinically available MRSI approaches. Due to the increased signal-to-noise ratio (SNR) and spectral resolution at higher field strengths, MRSI benefits from the use of modern ultra-high-field 7T systems. For example, it is possible to separate glutamate (Glu) and Gln at 7T, which is difficult at 3T due to the spectral overlap of the resonances of these metabolites [4].

**Magnetic Resonance Fingerprinting**

MRF, on the other hand, is a modern approach to mapping magnetic tissue properties, such as the T1 and T2 relaxation times [5]. Unlike conventional T1 and T2 mapping sequences, MRF derives the parameters of interest from a single acquisition wherein the flip angle, the repetition time (TR), and the echo time (TE) are varied pseudo-randomly during the acquisition of heavily undersampled data. The resulting data "fingerprint" can then be compared to a database, yielding T1 and T2 values. Since the result of this procedure is an actual T1 or T2 map and not just a T1- or T2- weighted image, MRF is considered a quantitative methodology, as it quantitatively estimates real physical quantities rather than providing arbitrary intensity parameters, which are more useful as a basis for machine-learning models. Similar to CRT-MRSI, MRF uses non-Cartesian k-space sampling to improve upon conventional T1 and T2 mapping sequences by minimizing the total acquisition duration.

**Motivation and Purpose**

After comparing the results obtained from MRSI acquisitions to those of clinical positron emission tomography (PET) scans as previously reported [2], we aimed to investigate the correlation between MRSI and MRF in glioma patients in this study, focusing on the correspondence between the hotspots identified in both methods. This constitutes an initial comparison of 7T MRSI and 3T MRF to determine whether the methods complement each other or whether they correlate.

The purpose of this work was to investigate, for the first time, whether metabolic changes detected by 7T MRSI correspond to structural changes found by 3T MRF in glioma patients by correlating the metabolic ratios of MRSI to T1 and T2 maps of MRF.

# Methods

## Study Population

We acquired approval of the institutional review board of the Medical University of Vienna (protocol 1991/2018), as well as written, informed consent from all participants of this prospective study. Participants were selected consecutively between February and December 2019. Inclusion criteria were a suspected glioma diagnosis, as well as informed consent, and the absence of MRI contraindications. Subjects were excluded if they were not eligible for a 7T MRI, if the MRSI data quality was too poor to allow reasonable data analysis, or if the subject's tumor could not be histologically confirmed as a glioma.

Patient recruitment is illustrated in Figure 1, and the cohort, consisting of 12 subjects (five females, seven males), 48±15 years of age, is listed in Table 1. There were two IDH-mutant grade 2 astrocytomas, three IDH-mutant grade 3 astrocytomas, two IDH-mutant grade 2 oligodendrogliomas, one IDH-mutant grade 3 oligodendroglioma, and four IDH-wildtype grade 4 glioblastomas, according to the 2021 WHO classification of gliomas [6]. The patient cohort in this paper overlapped with a cohort in previous papers (see Supplementary Table 3) [1,2].

## MRSI Protocol and Data Processing

The MRSI protocol was performed on a 7T Magnetom scanner (Siemens Healthineers) using a 1 Tx/32 Rx head coil (Nova Medical) and consisted of a T1-weighted MP2RAGE as the morphological reference, a B0 field map, a B1 field map for flip-angle optimization, and a CRT-FID-MRSI scan (TR = 450 ms, acquisition delay AD = 1.3 ms; FOV = 220 x 220 x 133 mm$^3$, resolution = 3.4 x 3.4 x 3.4 mm$^3$, TA = 15 min) [1,3].

MRSI post-processing used our previously introduced in-house pipeline and involved quantification in the spectral range of 1.8-4.1 ppm using LC Model [7]. A metabolite basis set consisting of 17 metabolites and a measured macromolecular baseline was used for fitting [8,9]. The metabolites included the previously mentioned neuro-oncological markers Cho (glycero- phosphocholine and phosphocholine, summarized as total choline, tCho), Cr (creatine and phospho-creatine, summarized as total creatine, tCr), Gln, Gly, Ins, and NAA (NAA together with NAA-glutamate, summarized as total NAA, tNAA), as well as γ-aminobutyric acid, glutathione, scyllo-inositol, serine, taurine, 2-hydroxyglutarate, and glutamate. An overview of the processing parameters is given in Supplementary Table 1 [10].

Data analysis included the ratios of tCho/tNAA, Gln/tNAA, Gly/tNAA, Ins/tNAA, tCho/tCr, Gln/tCr, Gly/tCr and Ins/tCr, as they are commonly used [2,11,12]. We specifically focused on the metabolite ratios to NAA because a drop in NAA, which corresponds to neuronal losses and is commonly seen in tumors, synergizes well with increases in tCho, Gln, Gly, and Ins, often producing well-defined hotspots in the ratio maps.

## MRF and Clinical Protocol

The MRF scan was performed on a 3T Magnetom PrismaFit MR scanner using a 1 Tx/64 Rx head coil (Siemens Healthineers), and was based on a 2D Fast Imaging with Steady-state Precession (FISP) spiral readout (FOV = 256x256 mm, in-plane resolution = 1x1 mm, TA = 20 s per slice). To reduce the MRF's long acquisition duration to an acceptable time, the number of acquired slices was kept as low as possible while still covering the entire tumor.

In addition to the MRF and MRSI protocol, a clinical routine 3T MRI was performed, consisting of a native T1-weighted image, a contrast-enhanced T1-weighted image and a fluid-suppressed T2-weighted image. The clinical images were segmented by a neuroradiologist. Co-registered clinical morphological scans and segmentations were used to define regions of interest (ROIs) according to different tissue characteristics: We distinguished between tumors with and without contrast uptake, and between necrotic and peritumoral tissue. The latter was defined by dilating a tumor mask by six voxels and subtracting the original tumor mask, resulting in an approximately 2 cm thick layer surrounding the tumor. We then analyzed different segmentations, namely, the tumor segmentation ("TU"), which included the tumor (including contrast-enhancing and non-contrast-enhancing tissue, necrosis, and edema), the dilated tumor segmentation ("TU+PT"), which added the peritumoral region, and the peritumoral segmentation alone ("PT").

**Data Analysis**

We compared these segmentations to a normal-appearing white matter (NAWM) reference region, which was created by subtracting TU+PT from a white matter mask and eroding the resulting region once. Additionally, we investigated metabolic abnormalities given by the median metabolite ratios and relaxation times in the different ROIs. Within each segmentation, we defined hotspots by including all voxels with a value greater than 150% of the respective median value of the NAWM reference region. For analysis, we compared the resulting T1 and T2 hotspots with the metabolite hotspots and with the tumor segmentation using Sørensen-Dice Similarity Coefficients (DSC), analogous to a previously established approach [2].

$$\text{DSC} = \frac{2 \times |N_{\text{MRSI}} \cap N_{\text{MRF}}|}{|N_{\text{MRSI}}| + |N_{\text{MRF}}|}$$

Since DSCs measure only the overlap of two regions, we also calculated the centers of intensity (i.e., the average position of all points of the ROI) of each region, according to

$$\vec{r}_{\text{VOI}} = \frac{\sum_{i \in \text{VOI}} \vec{v}_i \times I(\vec{v}_i)}{\sum_{i \in \text{VOI}} I(\vec{v}_i)},$$

with the voxel vectors $\vec{v}_i$ and the intensities $I(\vec{v}_i)$, and then evaluated their distances from each other ("center of intensity distances," COIDs):

$$\text{COID} = |\vec{r}_{\text{MRSI}} - \vec{r}_{\text{MRF}}|$$

Due to the possible tumor infiltrations of the surrounding regions, we extended our analysis to the peritumoral regions, again looking at DSCs between the MRF's T1 and T2 hotspots and MRSI's metabolic hotspots. In addition to the similarity measures, we evaluated median relaxation times and metabolic ratios in the hotspots within the different regions of interest (TU, TU+PT, PT) and the NAWM reference region. Last, we compared TU and PT using a two-sided paired Student's t-test. Since our approach of using a threshold to define the hotspots in TU and PT naturally increased the median values in these regions compared to the un-thresholded regions, the comparison to NAWM would have been meaningless and was thus omitted.

# Results

Overall, we found very high correspondence between the hotspots in the ratio maps for both Gln/tNAA and Gly/tNAA and the MRF's T1 and T2 maps, as well as the tumor segmentation, which is reflected in the respective DSCs and COIDs (see Figure 2 and Table 2).

**Median Relaxation Times and Metabolic Ratios**

Regarding the metabolic ratio values, the cohort's median in the tumor hotspot was highest for Ins/tNAA (median=1.15, [Q1, Q3] = [1.04, 1.21]), followed by Gln/tNAA (0.61, [0.56, 0.70]), tCho/tNAA (0.48, [0.42, 0.55]), and Gly/tNAA (0.28, [0.20, 0.36]), and the respective relaxation times were T1=1724 ms (Q1=1690 ms, Q3=1804 ms) and T2=85 ms (Q1=80 ms, Q3=106 ms). The corresponding values in NAWM were 0.54, [0.51, 0.59] for Ins/tNAA, 0.16, [0.13, 0.20] for Gln/tNAA, 0.20, [0.18, 0.21] for tCho/tNAA, and 0.07, [0.06, 0.10] for Gly/tNAA, and the relaxation times were T1=950 ms (Q1=941 ms, Q3=972 ms) and T2=42.9 ms (Q1=42.6 ms, Q3=43.3 ms).

Figure 3 shows an overview of the medians of metabolite ratios in the tumor hotspot while illustrating the different tumor grades by color-coding. The cohort's median metabolite ratios and median relaxation times for the hotspots in the TU, the PT, and the NAWM, are noted in Table 3 and shown in more detail in Figures 4 and 5. Notably, we found statistically significant differences between TU and PT in the metabolite ratios (with the values in TU higher than in PT), but no such effect was found for the relaxation times.

**Similarity Measures**

When comparing the hotspots of MRSI within the tumor to the entire segmentation TU, we found the highest DSC for Gln/tNAA (median = 0.78, [Q1, Q3] = [0.60, 0.91]), followed by Gly/tNAA (0.54, [0.48, 0.69]), tCho/tNAA (0.45, [0.35, 0.71]), and Ins/tNAA (0.35, [0.26, 0.53]). The DSCs for MRF were similar for both T1 (0.73, [0.66, 0.83]) and T2 (0.79, [0.67, 0.86]).

Comparing MRSI to the MRF's T1 hotspots in the tumor yielded the highest DSCs for Gln/tNAA (0.75, [0.54, 0.87]) and tCho/tNAA (0.61, [0.40, 0.73]), followed by Gly/tNAA (0.57, [0.46, 0.70]) and Ins/tNAA (0.43, [0.33, 0.52]). For T2, the DSCs were highest for Gln/tNAA (0.80, [0.68, 0.87]) and Gly/tnAA (0.62, [0.51, 0.73]), followed by tCho/tNAA (0.58, [0.47, 0.72]) and Ins/tNAA (0.41, [0.36, 0.53]). These results, together with the analogous results for the PT region, are noted in Table 2, and barplots of the entire cohort's tumor DSCs are displayed in Supplementary Figure 1.

The centers of intensity compared to the T1 hotspot were closest for Gln/tNAA (COIDS: median = 0.43 cm, [Q1, Q2] = [0.16 cm, 0.47 cm]) and Gly/tNAA (0.43, [0.29, 0.57]), and a bit higher for tCho/tNAA (0.48, [0.37, 0.59]) and Ins/tNAA (0.50, [0.44, 0.81]). For the T2 hotspot, the lowest COIDs were found for Gln/tNAA (0.21, [0.13, 0.33]) and Gly/tnAA (0.36, [0.19, 0.47]), and the values were again higher for tCho/tNAA (0.58, [0.34, 0.67]) and Ins/tNAA (0.58, [0.42, 0.73]). These values are illustrated in Figure 2.

**Complementary Information**

An example case is shown in Figure 6 in the form of the dataset of one selected patient with an IDH-mutant grade 3 astrocytoma, including the metabolic ratio maps of tCho/tNAA, Gln/tNAA, and Gly/tNAA, T1 and T2 maps from MRF, a T1w MP2RAGE, a T2w FLAIR (both acquired at 7T), and the radiologist's segmentation.

Last, Supplementary Figure 2 shows the median metabolic ratios for different ROIs for the threshold of 1.50 and illustrates the influence of the hotspot threshold on the median hotspot values. Part A of this figure notably shows the metabolites that exhibit the largest differences between TU and PT, and part B, by illustrating the case of a threshold value of 0.00, gives an indication of what the median values in the entire TU and PT regions (rather than the hotspot) would be.

## Discussion

We successfully conducted the first comparison of 7T MRSI and 3T MRF in 12 glioma patients, and found a high correspondence between the metabolic hotspots of Gln/tNAA and Gly/tNAA, the T1 and T2 relaxation time hotspots, and the radiologist's tumor segmentation, resulting in high DSCs and low COIDs for those two metabolite ratios, as shown in Table 2 and Figures 2 and 3. This finding complements our previous work [2], which showed a better correspondence of Gln/tNAA and Gly/tNAA to amino acid PET than the clinically used tumor marker tCho/tNAA.

Glutamine and glycine are amino acids that are involved in many metabolic processes in cells, including protein synthesis, energy production, and cell growth and repair [13,14]. For cancer cells, both glutamine and glycine can be the primary source of energy, and they also play a role in the proliferation of cancer cells [15]. Choline, on the other hand, is a polyatomic ion that plays an important role as a precursor of the phospholipid phosphatidylcholine, a major component of cell membranes, which is vital for their structural integrity and fluidity. Cancer cells tend to have a high demand for choline to sustain their proliferation [16].

Morphologically, the T1 and T2 relaxation times can change in tumors due to changes in the microenvironment of the tumor. For example, an accumulation of water in the cancer increases both the T1 and T2 times, as relaxation times in free water are longer than in bound water [17,18].

Our analysis of both MRF and MRSI data in the TU and PT segmentations showed similar T1 and T2 values, but significantly different metabolic ratios in the hotspots of both regions (Figures 3 and 4). The median values for metabolic ratios and relaxation times were much higher in these hotspots than in the NAWM control region due to the use of thresholding for hotspot definition.

Unfortunately, the existing literature on MRF in gliomas is still very limited [19]. De Blank et al. conducted MRF scans in a cohort of children and young adults with mostly low-grade gliomas, and found that T1 and T2 values tended to increase in tumors compared to a white matter control region. While there were some differences between their median values and ours in the tumor (T1: 1,444±254 ms, T2: 61±22 ms), the values in NAWM are comparable to those found in this study [20]. Springer et al. also found that T1 and T2 values from MRF increased in tumors compared to NAWM [21]. Regarding MRSI, the median values we found in this study are in accordance with our previous findings, which first suggested that Gln/tNAA and Gly/tNAA hotspots correspond well to PET in gliomas [2]. In addition, increases in tCho/tNAA have been commonly reported in the literature [22,23].

**Limitations and Outlook**

Due to the various types and grades of gliomas, as well as the small cohort size, it was not possible to separately analyze each tumor's diagnosis or grade. Due to the thresholding approach for hotspot definition, p-values could be calculated only to compare TU and PT, but not for the control region. Furthermore, MRSI is still an experimental modality and the quality of the results can vary between subjects from very good to unacceptably bad, necessitating the exclusion of some data sets. In addition, the availability of (clinical) 7T MRI systems is still limited, which, together with the rather long measurement times for MRF and MRSI, reduces the clinical applicability of this research for the time being. Last, the commonly used metabolic ratios present a weakness insofar as the overall ratio significantly depends on its denominator. Instead, concentration estimates offer more reliability and should be explored in future work.

Nevertheless, this preliminary study will provide a starting point for further studies, aiding in the development of more specific hypotheses which may be tested in larger cohort studies in the future, which should hopefully lead to better MRI-based delineation and classification of brain tumors. Ultimately, this work reinforces our previous finding that glutamine and glycine show great promise as potential biomarkers in glioma imaging via the use of ultra-high-field MR spectroscopy.

# Tables

| Cohort Overview |||||||
|---|---|---|---|---|---|
| Patient ID | Classification | Grade | IDH | Age | Sex |
| 1 | Glioblastoma | 4 | WT | 47 | F |
| 2 | Anaplastic Astrocytoma | 3 | Mut | 46 | F |
| 3 | Anaplastic Astrocytoma | 3 | Mut | 29 | M |
| 4 | Glioblastoma | 4 | WT | 52 | M |
| 5 | Diffuse Astrocytoma | 2 | Mut | 33 | M |
| 6 | Glioblastoma | 4 | WT | 58 | M |
| 7 | Diffuse Astrocytoma | 2 | Mut | 77 | F |
| 8 | Oligodendroglioma | 3 | Mut | 51 | M |
| 9 | Glioblastoma | 4 | WT | 61 | M |
| 10 | Anaplastic Astrocytoma | 3 | Mut | 28 | F |
| 11 | Oligodendroglioma | 2 | Mut | 38 | F |
| 12 | Oligodendroglioma | 2 | Mut | 61 | M |

Table 1: An overview of the cohort containing 12 patients, including the histological diagnosis according to the WHO 2021 classification, the tumor grade, the IDH1 mutation status (IDH1 mutant, *Mut*, and wild type, *WT*), the age at the time of the 7T MRSI measurement in years (average: 48±15), and the patient's sex (5 females, *F*, and 7 males, *M*).

| DSCs Between Different Hotspots | | | |
|---|---|---|---|
| Segmentations | TU | TU+PT | PT |
| DSC between | Median (Q1, Q3) | Median (Q1, Q3) | Median (Q1, Q3) |
| T1 & ROI | 0.73 (0.66, 0.83) | 0.47 (0.44, 0.52) | 0.58 (0.45, 0.65) |
| T2 & ROI | 0.79 (0.67, 0.86) | 0.46 (0.42, 0.54) | 0.58 (0.43, 0.62) |
| tCho/tNAA & ROI | 0.45 (0.35, 0.71) | 0.24 (0.16, 0.33) | 0.28 (0.13, 0.35) |
| Gln/tNAA & ROI | 0.78 (0.60, 0.91) | 0.55 (0.38, 0.59) | 0.65 (0.46, 0.80) |
| Gly/tNAA & ROI | 0.54 (0.48, 0.69) | 0.33 (0.28, 0.38) | 0.41 (0.34, 0.44) |
| Ins/tNAA & ROI | 0.35 (0.26, 0.53) | 0.21 (0.12, 0.23) | 0.25 (0.10, 0.28) |
| tCho/tNAA & T1 | 0.61 (0.40, 0.73) | 0.39 (0.25, 0.47) | 0.29 (0.13, 0.36) |
| Gln/tNAA & T1 | 0.75 (0.54, 0.87) | 0.60 (0.54, 0.64) | 0.51 (0.41, 0.56) |
| Gly/tNAA & T1 | 0.57 (0.46, 0.70) | 0.45 (0.38, 0.49) | 0.35 (0.31, 0.39) |
| Ins/tNAA & T1 | 0.43 (0.33, 0.52) | 0.32 (0.15, 0.37) | 0.25 (0.10, 0.30) |
| tCho/tNAA & T2 | 0.58 (0.47, 0.72) | 0.39 (0.26, 0.47) | 0.28 (0.14, 0.33) |
| Gln/tNAA & T2 | 0.80 (0.68, 0.87) | 0.61 (0.46, 0.64) | 0.47 (0.34, 0.56) |
| Gly/tNAA & T2 | 0.62 (0.51, 0.73) | 0.45 (0.39, 0.51) | 0.34 (0.29, 0.39) |
| Ins/tNAA & T2 | 0.41 (0.36, 0.53) | 0.33 (0.17, 0.38) | 0.25 (0.12, 0.29) |

Table 2: Sørensen-Dice Similarity Coefficients (DSCs, median and interquartile range IQR) of relaxation time hotspots (T1, T2), metabolite ratio hotspots (tCho/tNAA, Gln/tNAA, Gly/tNAA, Ins/tNAA), and different regions of interest (ROI), namely the tumor segmentation TU (containing non contrast-enhancing, contrast-enhancing, and necrotic tissue), the peritumoral region PT, and the combined region (TU+PT).

| Median Values in Different Regions of Interest | | | | |
|---|---|---|---|---|
| Segmentations | TU | TU+PT | PT | NAWM |
| Quantity | Median (Q1, Q3) | Median (Q1, Q3) | Median (Q1, Q3) | Median (Q1, Q3) |
| T1 | 1724 (1690, 1804) | 1770 (1712, 1792) | 1756 (1661, 1810) | 950 (941, 972) |
| T2 | 85.5 (80.1, 105.8) | 101.6 (94.0, 106.0) | 102.0 (90.0, 117.3) | 42.9 (42.6, 43.3) |
| tCho/tNAA | 0.48 (0.42, 0.55) | 0.40 (0.39, 0.49) | 0.38 (0.34, 0.44) | 0.20 (0.18, 0.21) |
| Gln/tNAA | 0.61 (0.56, 0.70) | 0.43 (0.40, 0.50) | 0.38 (0.35, 0.52) | 0.16 (0.13, 0.20) |
| Gly/tNAA | 0.28 (0.20, 0.36) | 0.22 (0.18, 0.26) | 0.20 (0.16, 0.24) | 0.07 (0.06, 0.10) |
| Ins/tNAA | 1.15 (1.04, 1.21) | 1.09 (0.94, 1.14) | 1.06 (0.90, 1.13) | 0.54 (0.51, 0.59) |
| tCho/tCr | 0.69 (0.63, 0.80) | 0.72 (0.65, 0.81) | 0.76 (0.66, 0.83) | 0.37 (0.35, 0.40) |
| Gln/tCr | 0.90 (0.75, 1.31) | 0.70 (0.61, 1.31) | 0.67 (0.58, 1.12) | 0.32 (0.26, 0.45) |
| Gly/tCr | 0.51 (0.33, 0.64) | 0.42 (0.30, 0.55) | 0.42 (0.28, 0.52) | 0.15 (0.10, 0.22) |
| Ins/tCr | 1.98 (1.88, 2.53) | 1.95 (1.85, 2.21) | 1.97 (1.85, 2.34) | 1.05 (0.96, 1.16) |

Table 3: Median values and first and third quartile (Q1, Q3) for the cohort's median T1 and T2 relaxation times and metabolic ratios in the hotspots within the tumor (TU), the peritumoral region (PT), and the p-values when comparing TU and PT using a two-sided paired Student's t-test, as well as the median values in the hotspots in the combined TU+PT region and in the normal appearing white matter control region (NAWM). P-values below 0.01 are in bold type. Notably, some metabolic ratios, such as Gln/tNAA, showed a statistically significant difference between TU and PT (e.g. $p < 0.001$ for Gln/tNAA), but the respective T1 and T2 values for MRF were not statistically significant. For a visualization of this data, see also Figures 3 and 4.

# Figures

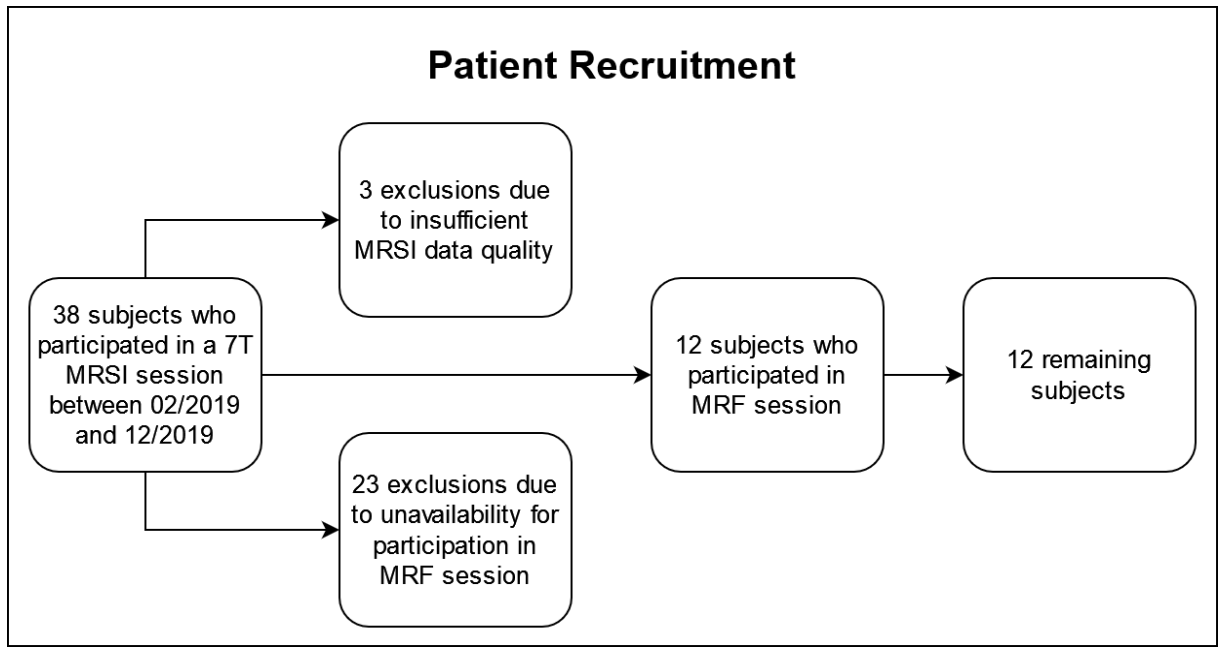

Figure 1: Flowchart of the recruitment for this study.

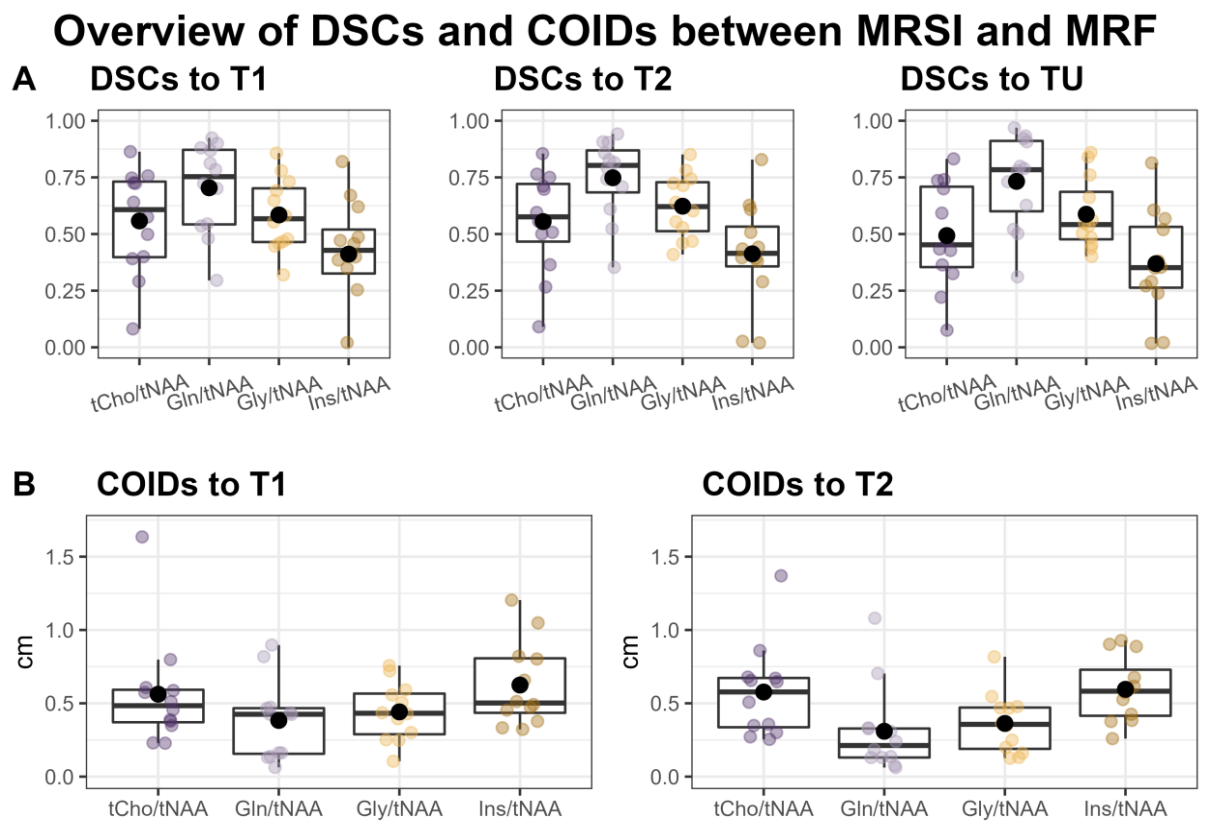

Figure 2: Overview of Sørensen-Dice Similarity Coefficients (DSCs) between the MRSI's metabolite ratio hotspots, MRF's T1 and T2 hotspots and the tumor segmentation (TU) (**A**) and the distances of the centers of intensity (COIDs) between MRSI and MRF hotspots (**B**).

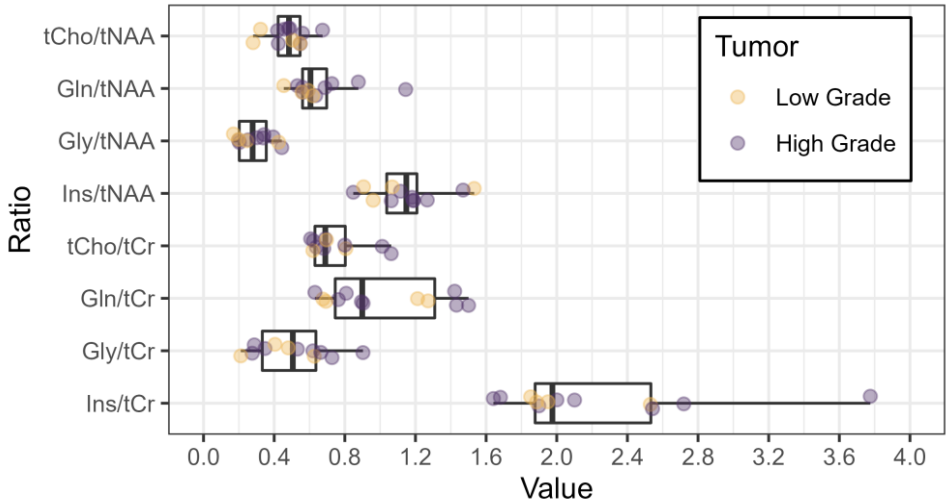

Figure 3: Boxplots of median metabolite ratios for tCho/tNAA, Gln/tNAA, Gly/tNAA, and Ins/tNAA, as well as tCho/tCr, Gln/tCr, Gly/tCr, and Ins/tCr, within the hotspot in the tumor segmentation TU. The colors indicate the tumor grade (yellow: low grade, grade 2 or below; violet: high grade, grade 3 or above).

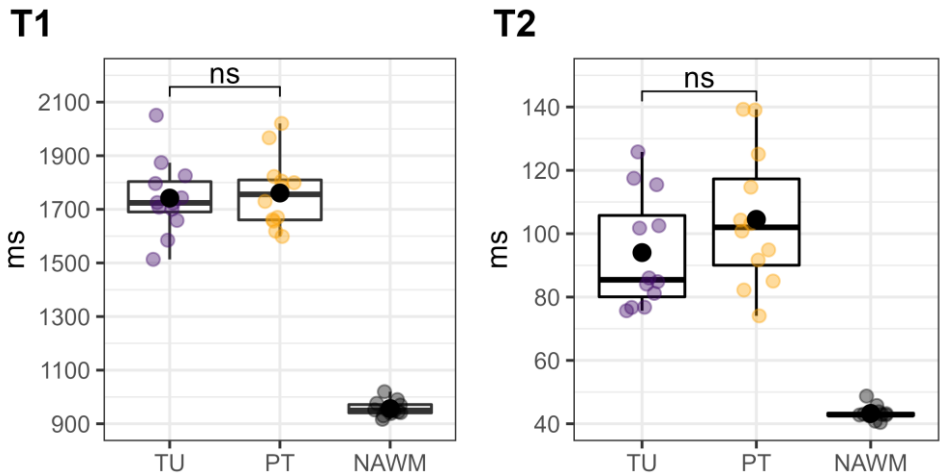

Figure 4: Median T1 (left) and T2 (right) relaxation times within the tumor (TU, violet) and peritumoral (PT, yellow) segmentations' hotspots, compared to the normal-appearing white matter (NAWM) control region (black). Each dot corresponds to one patient. TU and PT were compared using a two-sided paired t-test, which showed no significant difference ("ns"). Our approach of using a threshold to define the hotspots in TU and PT naturally increased the median values in these regions compared to the un-thresholded regions, which would have rendered a comparison to NAWM meaningless.

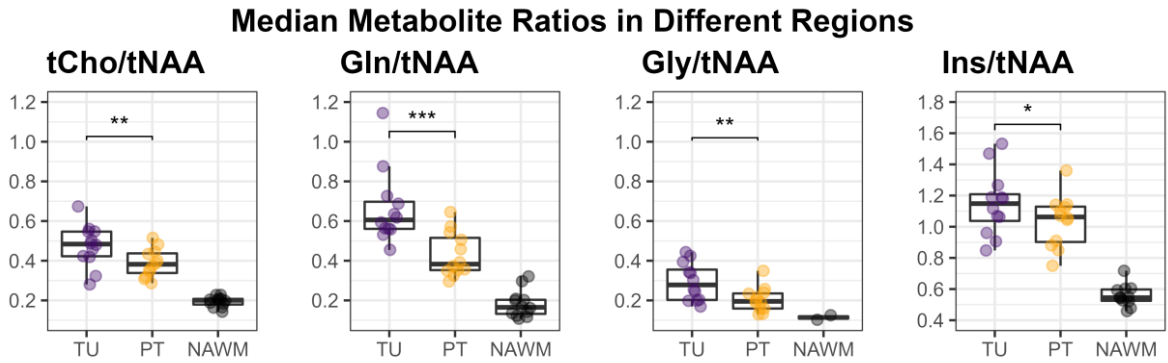

Figure 5: Median values for the metabolite ratios tCho/tNAA, Gln/tNAA, Gly/tNAA, and Ins/tNAA within the defined hotspots in the tumor (TU, violet) and peritumoral regions (PT, yellow), as well as the normal-appearing white matter control region (NAWM, black). The medians in the regions TU and PT were compared using a two-sided paired t-test and the resulting significance levels were noted in the plot (*: p<0.05, **: p<0.01, ***: p<0.001).

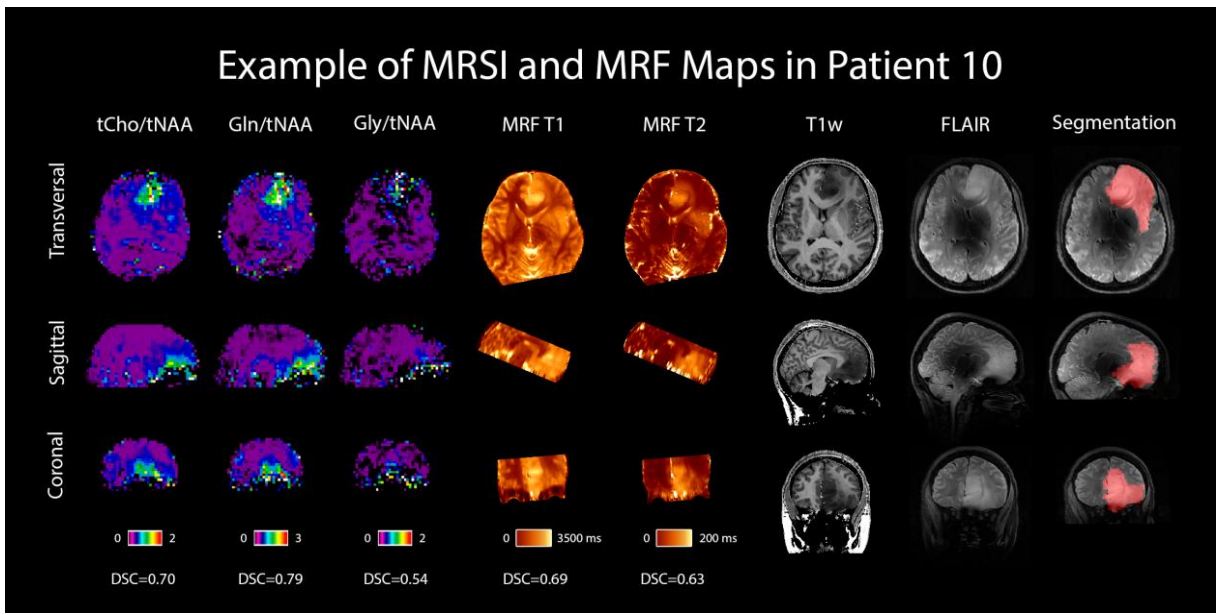

Figure 6: MRSI and MRF maps in a 28-year-old female patient with a histologically confirmed grade 3 astrocytoma. For comparison, 7T T1w MP2RAGE and FLAIR images are shown, as well as the radiologist's segmentation. Transversal, sagittal, and coronal views are shown, and Sørensen-Dice Similarity Coefficients comparing the hotspots to the segmentation are listed below the respective maps.

# Supplementary Material

| Minimum Reporting Standards in MR Spectroscopy - Overview | |
|---|---|
| Site | Vienna High Field MR Center |
| **1. Hardware** | |
| a. Field strength | 7T |
| b. Manufacturer | Siemens |
| c. Model | Magnetom |
| d. RF coils: nuclei (transmit/ receive), number of channels, type, body part | 1H, 32 ch, head, Nova Medical |
| e. Additional hardware | N/A |
| **2. Acquisition** | |
| a. Pulse sequence | FID-MRSI |
| b. Volume of interest (VOI) locations | Tumor, peritumoral, NAWM |
| c. Nominal VOI size | 220×220×110 mm³ |
| d. Repetition time (TR), echo time (TE) | 450 ms / 1.3 ms acquisition delay |
| e. Total number of excitations or acquisitions per spectrum | N/A, spatial-spectral encoding |
| In-time series for kinetic studies | N/A |
| i. Number of averaged spectra (NA) per time-point | N/A |
| ii. Averaging method (e.g., block-wise or moving average) | N/A |
| iii. Total number of spectra (acquired / in-time series) | N/A |
| f. Additional sequence parameters (spectral width in Hz, number of spectral points, frequency offsets); If STEAM: Mixing Time TM; If MRSI: 2D or 3D, FOV in all directions, matrix size, acceleration factors, sampling method | BW 2778 Hz, 1920 spectral points, MRSI: 3D, 220×220×133 mm³, 64×64×39, spatial-spectral encoding |
| g. Water suppression method | WET |
| h. Shimming method, reference peak, and thresholds for "acceptance of shim" chosen | Standard shim + manual adjustment, water peak < 50 Hz |
| i. Triggering or motion correction method | N/A |
| **3. Data analysis methods and outputs** | |
| a. Analysis software | LCModel 6.3-1 |
| b. Processing steps deviating from quoted reference or product | N/A |
| c. Output measure | Ratio |
| d. Quantification references and assumptions, fitting model assumptions | Simulated in NMRScope-B, macromolecular background |
| **4. Data Quality** | |
| a. Reported variables (SNR, linewidth (with ref. peaks)) | SNR and linewidths not reported |
| b. Data exclusion criteria | tCr SNR <5; tCr FWHM >0.15 ppm; metabolite Cramér-Rao lower bounds (CRLB) >40 % |
| c. Quality measures of post processing model fitting | CRLB |
| d. Sample spectrum | See Supp. Fig. 4 |

Supplementary Table 1: An overview according to the minimum reporting standards in MR spectroscopy [10].

| MRF Sequence Parameters | |
|---|---|
| Voxel Dimensions | 1.0 × 1.0 mm² |
| Matrix Size | 256 × 256 |
| Field of View | 256 × 256 mm² |
| Number of Slices | 10–13 |
| Slice Thickness | 5 mm |
| TE | 2 ms |
| TI | 21 ms |
| TR | 12.14–15.00 ms |
| TA | 3:51–4:51 (min:sec) |
| Acceleration Factor | 24 (inner k-space) |
| | 48 (outer k-space) |
| RX Bandwidth | 400 kHz |
| FA | 0–74° |
| Fat Saturation | no |

Supplementary Table 2: MRF sequence parameters.

| Patient ID | | |
|---|---|---|
| This Study | Hangel 2020 | Hangel 2022 |
| 1 | 3 | - |
| 2 | 4 | 2 |
| 3 | 5 | 3 |
| 4 | 6 | 4 |
| 5 | - | 5 |
| 6 | - | - |
| 7 | - | 6 |
| 8 | 10 | 8 |
| 9 | 14 | - |
| 10 | 17 | 13 |
| 11 | - | 15 |
| 12 | - | 16 |

Supplementary Table 3: Cohort overlap with previous publications [1,2].

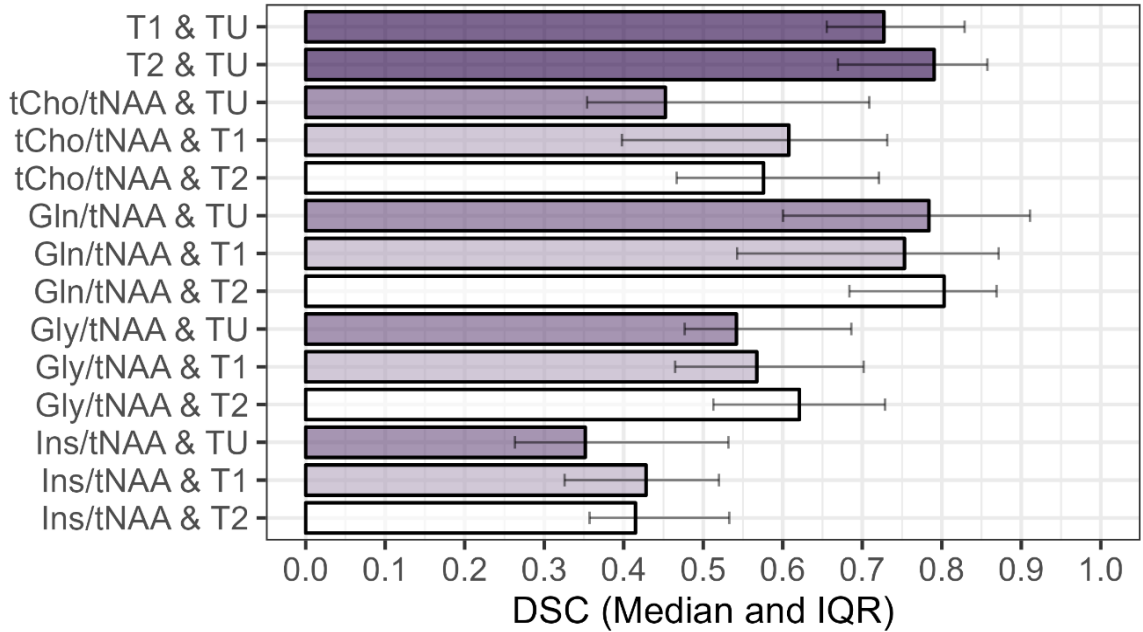

Supplementary Figure 1: Sørensen-Dice Similarity Coefficients (DSCs) in the tumor segmentation between the segmentation (TU), the MRF's relaxation time hotspots (T1, T2), and MRSI's metabolite ratios (tCho/tNAA, Gln/tNAA, Gly/tNAA, Ins/tNAA). Gln/tNAA had the highest correspondence with the tumor segmentation and T1 and T2 hotspots.

**Median Metabolite Ratios for Different Hotspot Thresholds**

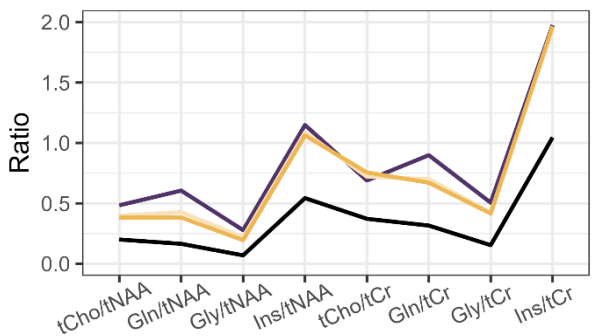
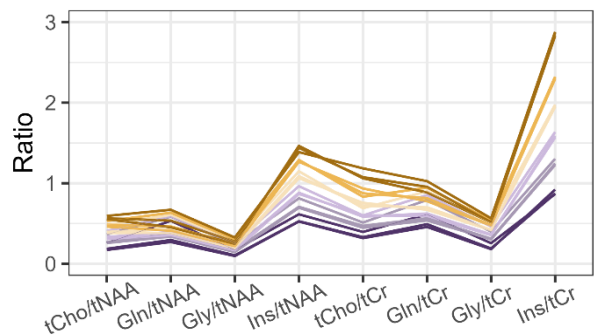
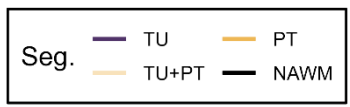
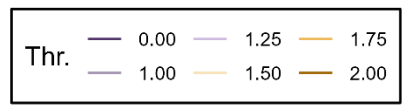

Supplementary Figure 2: **A**: Metabolite ratios tCho/tNAA, Gln/tNAA, Gly/tNAA, and Ins/tNAA, as well as tCho/tCr, Gln/tCr, Gly/tCr, and Ins/tCr, for a hotspot threshold of 1.5, plotted separately for the tumor segmentation (TU), the tumor region and the peritumoral region (TU+PT), the peritumoral region alone (PT), and the normal-appearing white matter control region (NAWM). **B**: The same metabolite ratios for different hotspot thresholds, with one line each for TU, TU+PT, and PT.

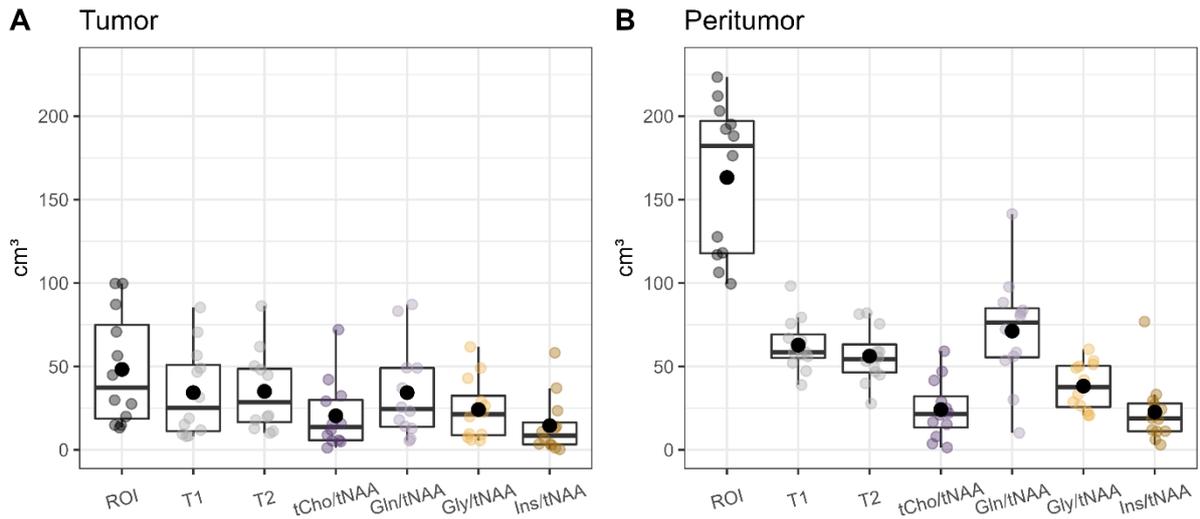

Supplementary Figure 3: Overview of the volumes of the region of interest (ROI), the MRF hotspots (T1, T2), and the MRSI hotspots (tCho/tNAA, Gln/tNAA, Gly/tNAA, Ins/tNAA), in the tumor (**A**) and the peritumoral region (**B**).

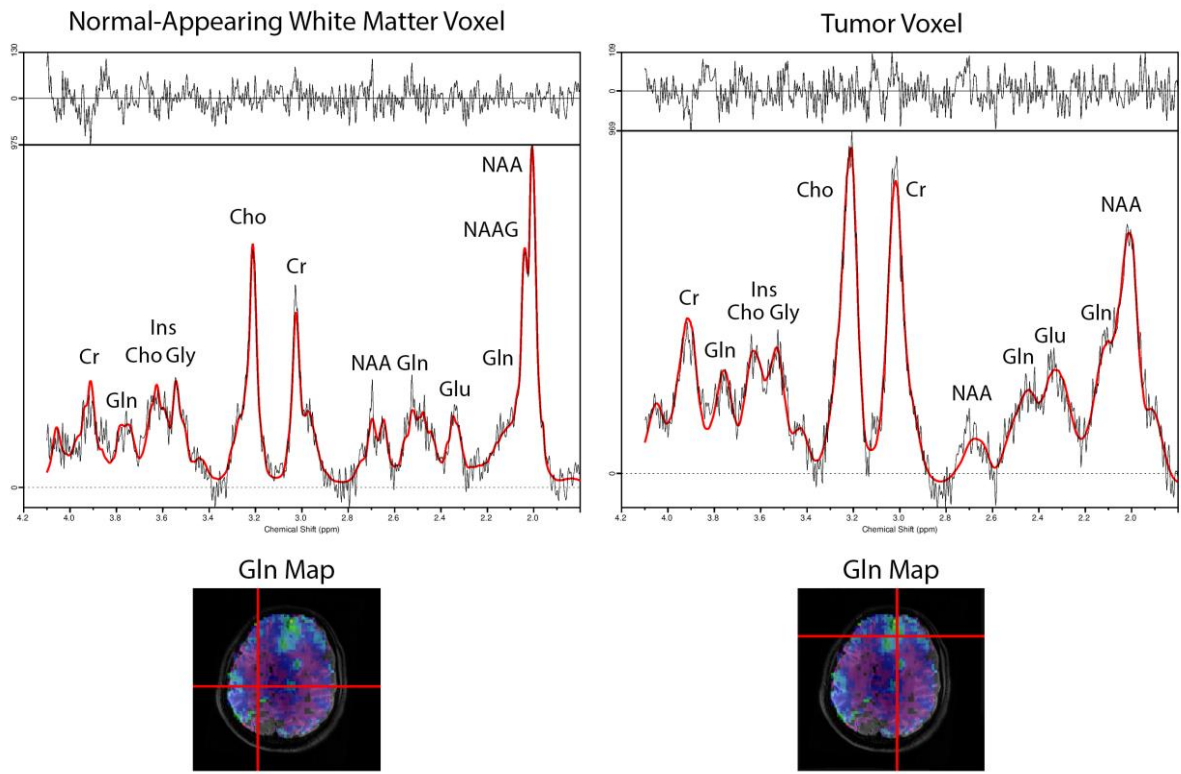

Supplementary Figure 4: Example spectra of patient 10 (anaplastic astrocytoma, grade 3, female, 28 years of age). Normal appearing white matter spectrum (left) and tumor spectrum (right). Below, glutamine (Gln) maps overlaid with a T1w reference image are shown, and the voxel position is indicated.

# Disclosures

1. *Acknowledgements*
   ChatGPT (OpenAI, San Francisco, CA) was used to aid in the creation of this manuscript.

2. *Funding*
   This work was supported by the Austrian Science Fund (FWF) projects KLI 646, KLI 1089, KLI 679. The financial support by the Austrian Federal Ministry for Digital and Economic Affairs and the National Foundation for Research, Technology and Development, and the Christian Doppler Research Association, as well as by the Comprehensive Cancer Center (Forschungsförderung der Initiative Krebsforschung, Medical University of Vienna) are gratefully acknowledged.

3. *Guarantor:*
   The scientific guarantor of this publication is Gilbert Hangel.

4. *Conflict of Interest:*
   The authors of this manuscript declare relationships with the following companies:
   - Julia Furtner has received honoraria for lectures and consultation from the following for-profit companies: Novartis; Seagen.
   - Matthias Preusser has received honoraria for lectures, consultation or advisory board participation from the following for-profit companies: Bayer; Bristol-Myers Squibb; Novartis; Gerson Lehrman Group (GLG); CMC Contrast; GlaxoSmithKline; Mundipharma; Roche; BMJ Journals; MedMedia; Astra Zeneca; AbbVie; Lilly; Medahead; Daiichi Sankyo; Sanofi; Merck Sharp & Dohme; Tocagen; Adastra; Gan & Lee Pharmaceuticals; Servier.

5. *Statistics and Biometry:*
   No complex statistical methods were necessary for this paper.

6. *Informed Consent:*
   Written informed consent was obtained from all subjects (patients) in this study.

7. *Ethical Approval:*
   Institutional Review Board (IBR) approval was obtained by the IBR of the Medical University of Vienna, protocol 1991/2018.

8. *Study subjects or cohorts overlap:*
   Some study subjects or cohorts have been previously reported in:
   - G. Hangel, C. Cadrien, P. Lazen, et al. (2020) High-resolution metabolic imaging of high-grade gliomas using 7T-CRT-FID-MRSI, NeuroImage Clin. DOI: 10.1016/j.nicl.2020.102433.
   - G. Hangel, P. Lazen, S. Sharma, et al. (2022) 7T HR FID-MRSI Compared to Amino Acid PET: Glutamine and Glycine as Promising Biomarkers in Brain Tumors, Cancers. DOI: 10.3390/cancers14092163.

   This study was the first comparison of MR spectroscopic imaging (MRSI) and MR fingerprinting in gliomas, and thus differs substantially from the previous publications, which covered 7T MRSI in gliomas in general (Hangel 2020) and a comparison of MRSI and PET in gliomas (Hangel 2022). Of the 12 patients included in this study, 7 were included in Hangel 2020, and 9 were included in Hangel 2020. The exact overlap is noted in Supplementary Table 3.

9. *Methodology*
   Prospective cross-sectional study, performed at one institution.